\documentclass[traditabstract]{aa}  
\usepackage{graphicx}
\usepackage{txfonts}
\usepackage{natbib}
\bibpunct{(}{)}{;}{a}{}{,}	
\begin{document}

   \title{High-fidelity view of the structure and fragmentation of the high-mass, filamentary IRDC G11.11-0.12}

   \author{J. Kainulainen	\inst{1},
           S. E. Ragan 		\inst{1},  
           T. Henning		\inst{1}, \and 
           A. Stutz			\inst{1}
          }
   \offprints{jtkainul@mpia.de}

   \institute{Max-Planck-Institute for Astronomy, K\"onigstuhl 17, 69117 Heidelberg, Germany \\
              \email{jtkainul@mpia.de}
            }
   \date{Received ; accepted }
\abstract{Star formation in molecular clouds is intimately linked to their internal mass distribution.  
We present an unprecedentedly detailed analysis of the column density structure of a high-mass, filamentary molecular cloud, namely IRDC G11.11-0.12 (G11). We use two novel column density mapping techniques: high-resolution ($FWHM=2\arcsec$, or $\sim 0.035$ pc) dust extinction mapping in near- and mid-infrared, and dust emission mapping with the \emph{Herschel} satellite. These two completely independent techniques yield a strikingly good agreement, highlighting their complementarity and robustness. We first analyze the dense gas mass fraction and linear mass density of G11. We show that G11 has a top heavy mass distribution and has a linear mass density ($M_\mathrm{l} \sim 600$ M$_\odot$ pc$^{-1}$) that greatly exceeds the critical value of a self-gravitating, non-turbulent cylinder. These properties make G11 analogous to the Orion A cloud, despite its low star-forming activity. This suggests that the amount of dense gas in molecular clouds is more closely connected to environmental parameters or global processes than to the star-forming efficiency of the cloud. We then examine hierarchical fragmentation in G11 over a wide range of size-scales and densities. We show that at scales $ 0.5\ \mathrm{pc} \gtrsim l \gtrsim 8$ pc, the fragmentation of G11 is in agreement with that of a self-gravitating cylinder. At scales smaller than $l \lesssim 0.5$ pc, the results agree better with spherical Jeans' fragmentation. One possible explanation for the change in fragmentation characteristics is the size-scale-dependent collapse time-scale that results from the finite size of real molecular clouds: at scales $l \lesssim 0.5$ pc, fragmentation becomes sufficiently rapid to be unaffected by global instabilities.
}
   \keywords{ISM: clouds - ISM: structure - Stars: formation - dust, extinction} 
  \authorrunning{J. Kainulainen et al.}
  \titlerunning{High-fidelity view of G11.11-0.12}
  \maketitle


\section{Introduction} 
\label{sec:intro}


Measuring mass distributions of molecular clouds is of crucial importance for understanding processes that regulate star formation \citep[][]{hen12}. However, probing the entire wide range of (column) densities that molecular clouds show, at a resolution that resolves their substructure, is an observational challenge. This is especially true for high-mass clouds that potentially harbor precursors of high-mass stars because of their higher column densities and larger distances. Owing to these difficulties, a global description of the relation between molecular cloud structure and star formation is still elusive. 


The \emph{Herschel} satellite \citep{pil10} provides an outstanding tool to study the structure of high-mass molecular clouds at different scales \citep[e.g.,][]{beu10, sch12}, thanks to its high mapping speed and dynamic range. In particular, early studies employing \emph{Herschel} data have signified the role of \emph{filamentary} structures as an important pathway toward star formation \citep[e.g.,][]{and10}. However, \emph{Herschel} observations only provide column density data in a spatial resolution of $\sim 25\arcsec$ ($\sim 0.4$\,pc at 3.5\,kpc), while molecular clouds show fragmentation down to at least $\lesssim 0.03$\,pc \citep[e.g.,][]{sta06, per09}. 

We have developed an alternative, novel technique to map column densities in infrared dark clouds (IRDCs) using dust extinction measurements in near- and mid-infrared \citep[NIR and MIR,][]{kai13}. The technique takes advantage of the high resolution of the \emph{Spitzer} 8\,$\mu$m images ($FWHM = 2\arcsec$) and of the relatively good sensitivity of NIR-based extinction data \citep[cf.,][]{kai11alves}. When combined, these data can provide unique, high-resolution column density data that cover a relatively wide range of column densities, $N(\mathrm{H}_\mathrm{2}) \approx 2-150 \times 10^{21}$ cm$^{-2}$. Importantly, the assumptions on which the technique is based are completely different to those employed with dust emission techniques. This makes dust extinction data highly complementary to \emph{Herschel} data, in addition to the advantage that it provides more than ten times higher spatial resolution.


The IRDC G11.11-0.12 (G11) is an excellent target for studying the structure of a young, high-mass cloud at the onset of star formation. The $\sim$30 pc long filamentary cloud contains a large reservoir of cold gas as evidenced by its MIR and sub-mm properties \citep{car00, joh03, hen10}. It harbors 19 \emph{Herschel} point sources, most of which contain 24\,$\mu$m sources indicating the presence of a protostar \citep{hen10, rag12}. It also contains one possible high-mass star-forming core \citep{pil06a}.

In this paper, we present an analysis of the density structure of G11 using high-resolution ($FWHM = 2\arcsec$ corresponding to 0.035\,pc assuming $d=3.6$\,kpc) dust extinction data and \emph{Herschel} dust emission data. The data allow us to cover, for the first time in an IRDC, simultaneously the scales from tenth-of-a-parsec cores to the large-scale, tenuous envelope of the cloud.  

\section{Observations}           
\label{sec:observations}

\subsection{High-dynamic-range dust extinction data}  
\label{subsec:avdata}

We derived column density data for G11 using the extinction mapping technique presented by \citet{kai13}. The technique combines MIR (8\,$\mu$m) surface brightness data with NIR ($JHK$) photometry of stars. 
We derived an 8\,$\mu$m optical depth ($\tau_{8\,{\mu}m}$) map of G11 using the \emph{Spitzer/GLIMPSE} data  \citep{ben03} and the approach detailed in \citet{rag09}. Then, we derived a NIR extinction map using data from the \emph{UKIRT/Galactic Plane Survey} \citep{law07} and the procedure described in \citet{kai11alves}. The data sets were then combined as described in \citet{kai13}. For the procedure, it is necessary to adopt the relative dust opacity law between the $JHK$ bands and 8\,$\mu$m. We adopted the relations \citep[][]{car89, oss94}
\begin{equation}
\tau_\mathrm{K} = 0.60 \tau_\mathrm{H} = 0.40 \tau_\mathrm{J} = 3.4 \tau_{8\,\mu \mathrm{m}} = 0.11\tau_V.
\label{eq:dustlaw}
\end{equation}
Finally, the optical depths were used to estimate column densities via the conversion \citep{sav77, boh78}
\begin{equation}
N(\mathrm{H}_2) = 0.94\times 10^{21} \mathrm{\ cm}^{-2}~\big( \frac{A_\mathrm{V}}{\mathrm{mag}} \big).
\label{eq:bohlin}
\end{equation}
Figure \ref{fig:g11} shows the resulting column density map of G11. It has the spatial resolution of $2\arcsec$ and 3$-\sigma$ sensitivity of $N(\mathrm{H}_2)\approx 2 \times 10^{21}$ cm$^{-2}$. The zero-point uncertainty of the map is about $\sigma_{\mathrm{0}, N(\mathrm{H}_\mathrm{2})} \approx 1 \times 10^{21}$ cm$^{-2}$ \citep[cf.,][]{kai11alves}. 

\subsection{\emph{Herschel} data}                
\label{subsec:herschel}

We derived column densities in G11 using \emph{Herschel} data. The data were obtained under the \emph{Earliest Phases of Star formation} guaranteed time key programme (PI: O. Krause) and are described in \citet{rag12}. The data were processed to level 1 using HIPE \citep{ott10}, developer build 10.0(2538), calibration trees 42 (PACS) and 10.0 (SPIRE). We generated final maps using {\sc Scanamorphos} version 20 \citep{Roussel2012}, with the {\tt 'galactic'} option, and included the non-zero-acceleration telescope turn-around data.
Using the spectral energy distribution fitting method \citep{Launhardt2013}, we computed $N(\mathrm{H}_\mathrm{2})$ for positions with emission in at least four bands of the {\em Herschel} 70, 100, 160, 250, and $350\,\mu$m data and 870\,$\mu$m ATLASGAL data \citep{sch09}. All maps were first convolved to the resolution of the 350\,$\mu$m maps ($24\farcs9$) and re-gridded to match the $6\arcsec$-per-pixel SPIRE grid scale. We present the resulting column density and temperature maps in Ragan et al. (in prep.).

Since we later compare the \emph{Herschel}-derived column densities with the extinction-derived ones, it is important to ascertain that the dust opacities (that directly affect the column densities) are chosen consistently from sub-mm to NIR. Dust opacities are not very well constrained, as the range of plausible models is at least a factor of five in sub-mm and two in 8 $\mu$m and NIR (shown in Fig. \ref{fig:opacities}). The effective dust opacities used in calculating column densities from NIR and 8 $\mu$m \citep[see][]{kai13} are close to the means of the models shown in Fig. \ref{fig:opacities}. Therefore, we choose to use the mean values from the models also in calculating the column densities from  \emph{Herschel} data (see Fig. \ref{fig:opacities}). These values are formally closest to the \citet{oss94} dust model without dust coagulation.

\begin{figure*}
   \centering
\includegraphics[bb = 97 55 640 450, clip=true, angle=270, width=0.95\textwidth]{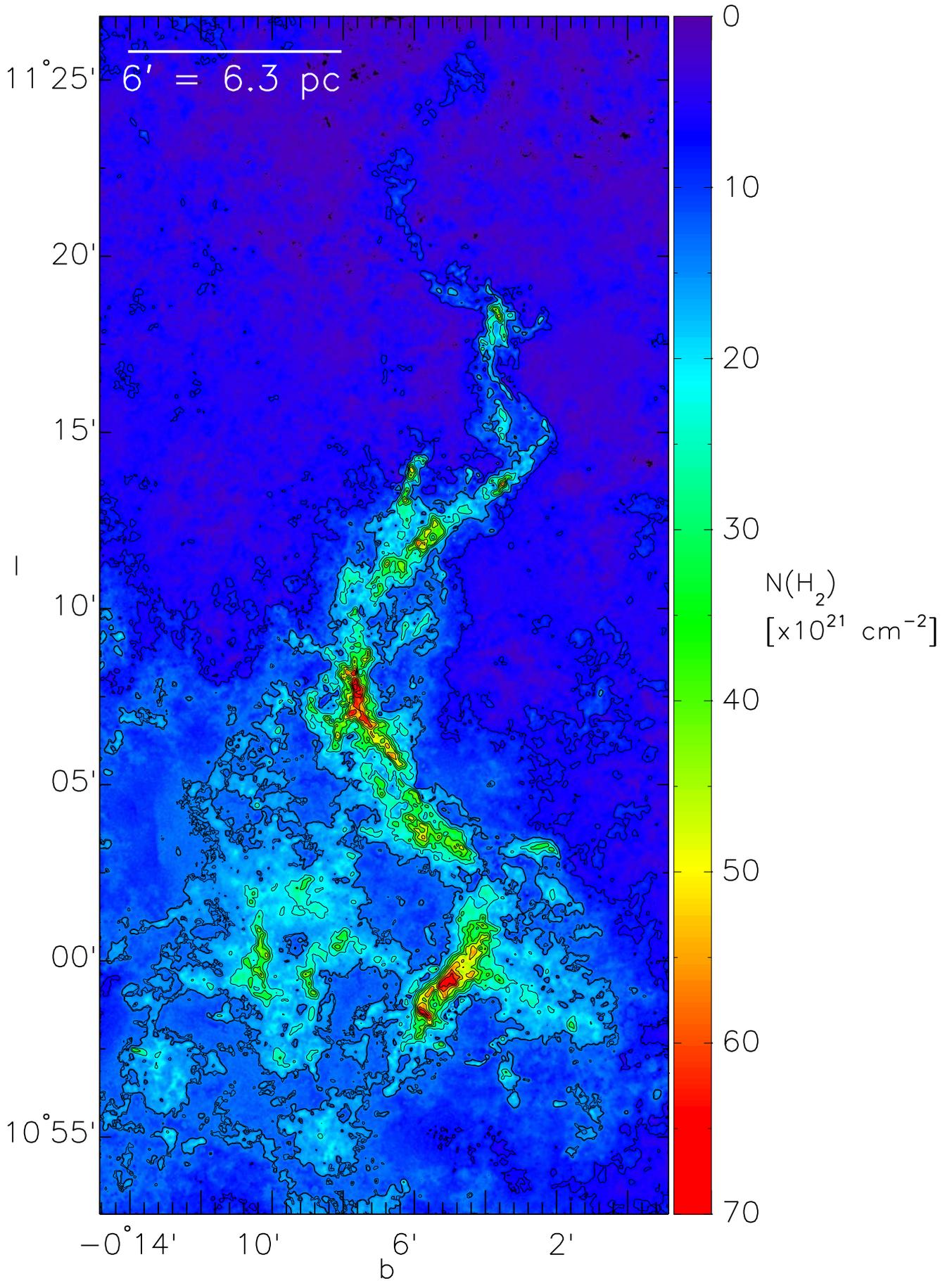}
      \caption{$N(\mathrm{H}_2)$ map of IRDC G11.11-0.12, derived using NIR and MIR dust extinction data. The spatial resolution of the map is $2\arcsec$ and it covers the dynamic range of $N_\mathrm{H_2} = 2-95 \times 10^{21}$ cm$^{-2}$. The contour levels are $[3, 10, 15, 25, \dots] \times 10^{21}$ cm$^{-2}$. The contour at $10 \times 10^{21}$ cm$^{-2}$ is bold. The coordinate system is tilted: the Galactic longitude is on y-axis and latitude on x-axis.
              }
         \label{fig:g11}
   \end{figure*}

The column density/temperature fitting technique we employ \citep{Launhardt2013} does not take into account that significant amount of the detected emission may result from the extended, diffuse Galactic dust component (i.e., not from the G11 cloud). This is because the \citet{Launhardt2013} work targets specifically objects \emph{off} the Galactic plane where the dust component is negligible. However, towards the Galactic plane where G11 resides, the total column density of the diffuse component can easily amount to several $\times 10^{21}$ cm$^{-2}$ \citep[e.g.,][]{mar06}. Also, spatial variations are expected. We performed a Monte Carlo simulation to test if the variable background component can result in systematic errors and/or increased uncertainty in the derived temperatures/column densities. We first estimated the level of possible background variations from the observed data by measuring the standard deviation, $\sigma_\mathrm{bg}(\lambda )$, of emission values from an extended area at each wavelength. For this area, we chose all pixels in which the 250 $\mu$m emission was below 15 mJy/arcsec$^2$. Qualitatively, the area corresponds about to the area below $N(\mathrm{H}_2) \lesssim 5 \times 10^{21}$ cm$^{-2}$ in the map we present in Fig. \ref{fig:g11}. The area is extended and likely contains dust that is related to G11. Therefore, our estimates should reflect the \emph{maximum} of possible random background variations. The standard deviations were $\sigma_\mathrm{bg}(\lambda ) = \{6, 16, 10, 5\}$ mJy/arcsec$^2$ for the $\{70, 160, 250, 350 \} \ \mu$m \emph{Herschel} bands, respectively. The ATLASGAL data employs, by construction, strong spatial filtering compared to \emph{Herschel} data, and possible background variations are efficiently filtered out. Then, we generated model SEDs and applied to each set a wavelength dependent, normally distributed systematic offset. The dispersion of the Gaussian distribution from which the offsets were drawn was set to the standard deviation measured previously from the data, i.e., to $\sigma_\mathrm{bg}(\lambda )$. The mean of the distribution was set to zero, and consequently, the simulation describes variations that average out over the entire field. Finally, we calculated the new temperature and column density using the ``perturbed'' SED. The procedure was repeated $10^3$ times for 14 model column densities between  $N(\mathrm{H}_2) = 1-200 \times 10^{21}$ cm$^{-2}$ and for three values of temperatures, namely $T = \{15, 20, 25\}$ K.

The results of the Monte Carlo experiment are summarized in Fig. \ref{fig:bg_sim} and show that the derived temperatures do not suffer from a systematic error because of possible random variations in the background emission. The uncertainty of the temperature measurements is increased, depending on the temperature of the diffuse dust component: for the model temperatures of $\{15, 20, 25\}$ K, the inter-quartile ranges of the derived temperatures are $\{1.7, 2.7, 0.7\}$ K, respectively. Thus, it appears that the uncertainty in the temperature measurement is increased by roughly 2 K if background variations are as strong as we estimate from the data. Similar effect can be seen in the derived column densities. There is no significant systematic error, but the uncertainty is increased. The inter-quartile ranges of the input/output column density ratios are about 15\% and standard deviations about 30\%, irrespective of the input column density. In conclusion, we show that random variations of the emission from the diffuse Galactic dust component do not cause systematic errors to the \emph{Herschel}-derived column densities but do contribute significantly to the random errors (noise). However, significant systematic gradients in the diffuse dust component that do not average out over the field could still affect the temperature/column density measurements. The large-scale gradients across the field in the NIR extinction data \citep{kai11alves} are roughly $N(\mathrm{H}_2) \approx 5 \times 10^{21}$ cm$^{-2}$. The NIR extinction data recover low-column densities relatively well and the NIR technique does not include spatial filtering. This implies that the possible effects due to systematic background variations should be limited to column densities that are clearly below $N(\mathrm{H}_2) \lesssim 10 \times 10^{21}$ cm$^{-2}$.

\section{Results and Discussion}           
\label{sec:results}


\subsection{Comparison of Herschel and dust extinction data}           
\label{subsec:comparison}

\begin{figure}
   \centering
\includegraphics[bb = 90 25 430 330, clip=true, width=0.5\textwidth]{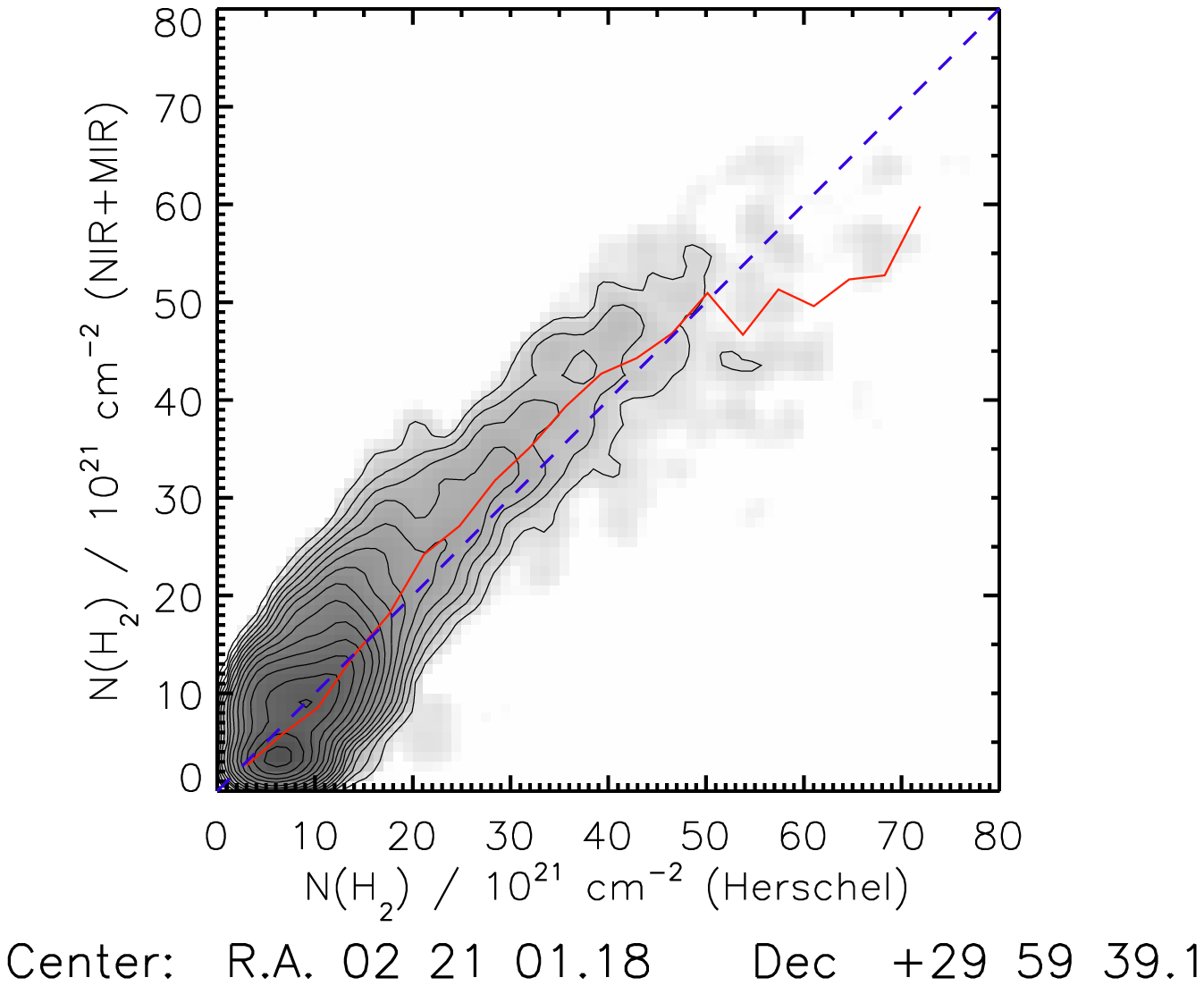}
\includegraphics[bb = 0 0 500 350, clip=true, width=0.45\textwidth]{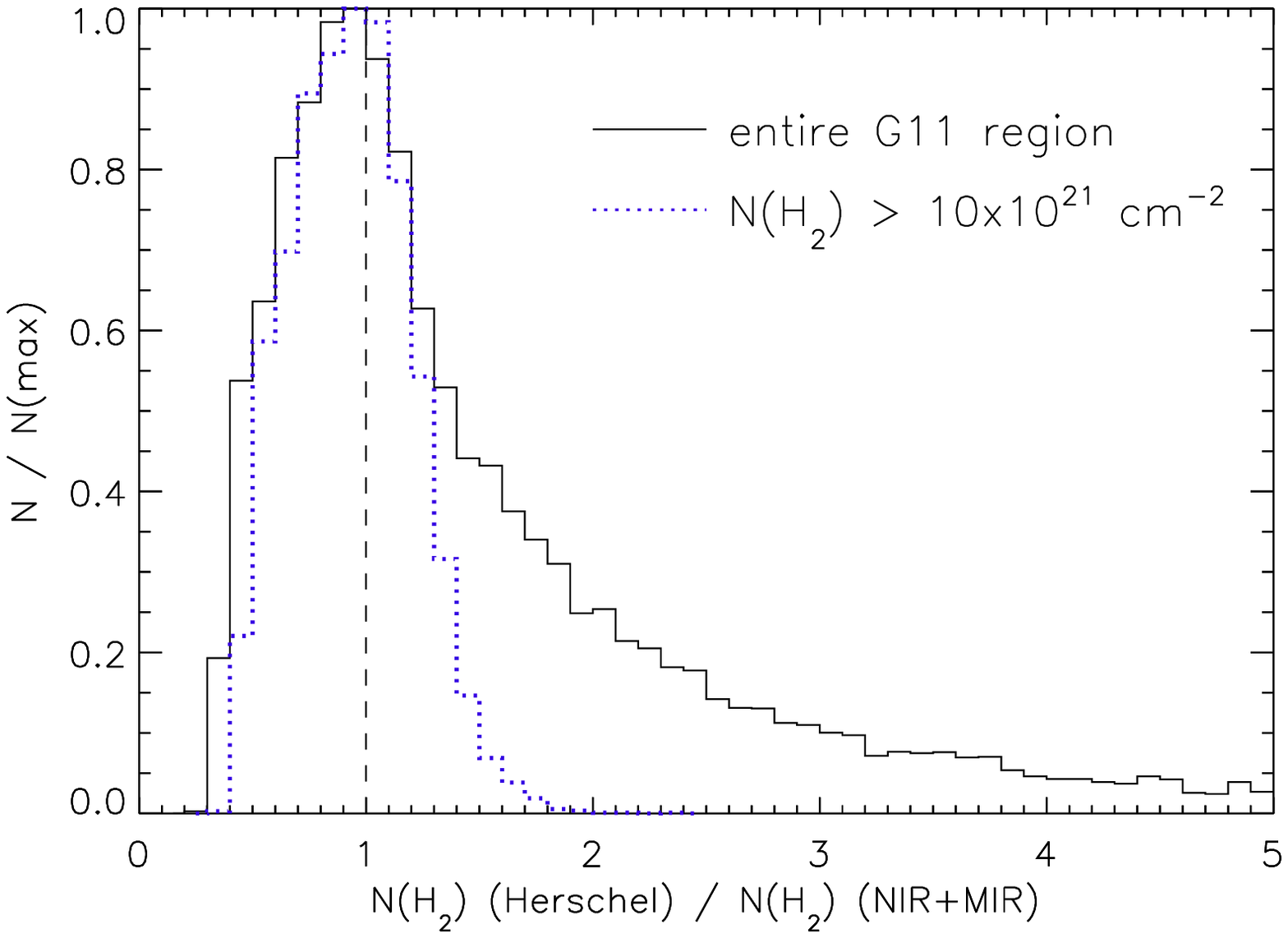}
      \caption{\emph{Top: }Pixel-to-pixel comparison of column densities derived with the NIR+MIR dust extinction technique and the Herschel dust emission mapping. The color scale and contours show the density of data points in the scatter plot. The dashed line shows the one-to-one relation. The red solid line shows the median relationship. \emph{Bottom: }Histogram of the ratio of Herschel-derived column densities to NIR+MIR dust extinction derived column densities. The black solid line shows the histogram for all the data in the field. The blue line shows the histogram for the area with column density over $N(\mathrm{H}_2) > 10 \times 10^{21}$ cm$^{-2}$. The dashed line shows the ratio of one.
              }
         \label{fig:cds}
   \end{figure}

We first examined how the dust extinction- and \emph{Herschel}-derived column densities compare by making a pixel-to-pixel comparison. 
For this analysis, the extinction data were smoothed to the $FWHM=24\farcs9$ \emph{Herschel} resolution. The comparison shows (Fig. \ref{fig:cds}, top frame) that the two column density measurement techniques are in a very good agreement. Note that the column densities (at this resolution) extend only to about $N(\mathrm{H}_2) \lesssim 70 \times 10^{21}$ cm$^{-2}$. Figure \ref{fig:cds} (bottom frame) shows a histogram of the ratio of the two column density measurements. The histogram is shown separately for the entire mapped area (black curve) and for the area with $N(\mathrm{H}_2) > 10 \times 10^{21}$ cm$^{-2}$. The histograms peak strongly at a ratio of unity. About $70\%$ of all pixel values are within a factor of two from one-to-one relation. Above $N(\mathrm{H}_2) > 10 \times 10^{21}$ cm$^{-2}$, the corresponding percentage is 90\%. The full histogram has a tail toward higher ratios (i.e., \emph{Herschel} data gives higher column densities) that can be attributed to a low-column density area $N(\mathrm{H}_2) \lesssim 5 \times 10^{21}$ cm$^{-2}$ in the north-west, part of the mapped region. It is possible that these discrepant values result, at least partly, from large-scale, systematic gradients in the emission detected from the diffuse Galactic dust component  (our \emph{Herschel} data analysis does not take such gradients into account). The histogram of the high-column density part does not show this discrepancy, indicating that if systematic background gradients are present, their effect is restricted to low-column density regions. 

We conclude that the \emph{column densities derived from dust extinction and Herschel are in excellent agreement, given our current knowledge of dust opacities}. As explained in Section \ref{subsec:herschel}, the dust opacities were chosen as a mean of plausible models, shown in Fig. \ref{fig:opacities}, which are nominally closest to the \cite{oss94} model without dust coagulation. It is an interesting topic for future studies to examine whether this result holds in a larger sample of IRDCs. It could be hypothesized that dust coagulation may significantly alter dust opacities at the column densities our maps cover. However, as we will show in Section \ref{subsec:fragmentation}, the mean volume densities at the scale of the Herschel beam ($\sim 0.4$ pc) are  about $10^4$ cm$^{-3}$, not expected to be high enough for coagulation to be efficient \citep[or, alternatively, would require relatively long time-scales, see, e.g.,][]{oss94}.


\subsection{Column density distribution of G11}           
\label{subsec:cd}


Figure \ref{fig:g11} shows the $N(\mathrm{H}_\mathrm{2})$ map of G11 derived from dust extinction. 
%
%
The total mass in the mapped area is $M_\mathrm{tot} \approx 10^5$ M$_\odot$, regardless whether estimated from dust extinction or \emph{Herschel} data. We estimate the mass associated with the filament-shaped part of the cloud from the column densities above $N(\mathrm{H}_\mathrm{2}) > 10 \times 10^{21}$ cm$^{-2}$, resulting in $M_\mathrm{fil} \approx [1.3, 2.5] \times 10^4$ M$_\odot$. The lower limit is the mass after a subtraction of constant $N(\mathrm{H}_\mathrm{2}) = 10 \times 10^{21}$ cm$^{-2}$ from the filament area (to mimic a surrounding envelope), and the upper limit is the mass without subtraction. We find mean linear mass densities of $M_\mathrm{l} = [430, 850]$ M$_\odot$ pc$^{-1}$ for the 30\,pc long filament, i.e., on average $M_\mathrm{l} \approx 600$ M$_\odot$ pc$^{-1}$. The critical value for a self-gravitating, \emph{non-turbulent} cylinder is $M_\mathrm{l}^\mathrm{crit} = 2 c_\mathrm{s}^2 / G \approx 25$ M$_\odot \big( \frac{T}{15 \ \mathrm{K}} \big)$, where $c_\mathrm{s}$ is the speed of sound \citep{ost64}. In the case of turbulent support, the condition is $M_\mathrm{l}^\mathrm{crit} = 84 (\Delta V)^2$ M$_\odot$ pc$^{-1}$ \citep[][]{jac10}. Assuming a typical value of $\Delta V = 2.5$ km s$^{-1}$ (see Section \ref{subsec:fragmentation}) results in $M_\mathrm{l} = 525$ M$_\odot$ pc$^{-1}$. The linear mass density of G11 greatly exceeds the non-turbulent critical value and is approximately in agreement with the turbulent one. G11 clearly must be dominated by non-thermal motions.

\begin{figure}
   \centering
\includegraphics[bb = 0 8 500 339, clip=true, width=\columnwidth]{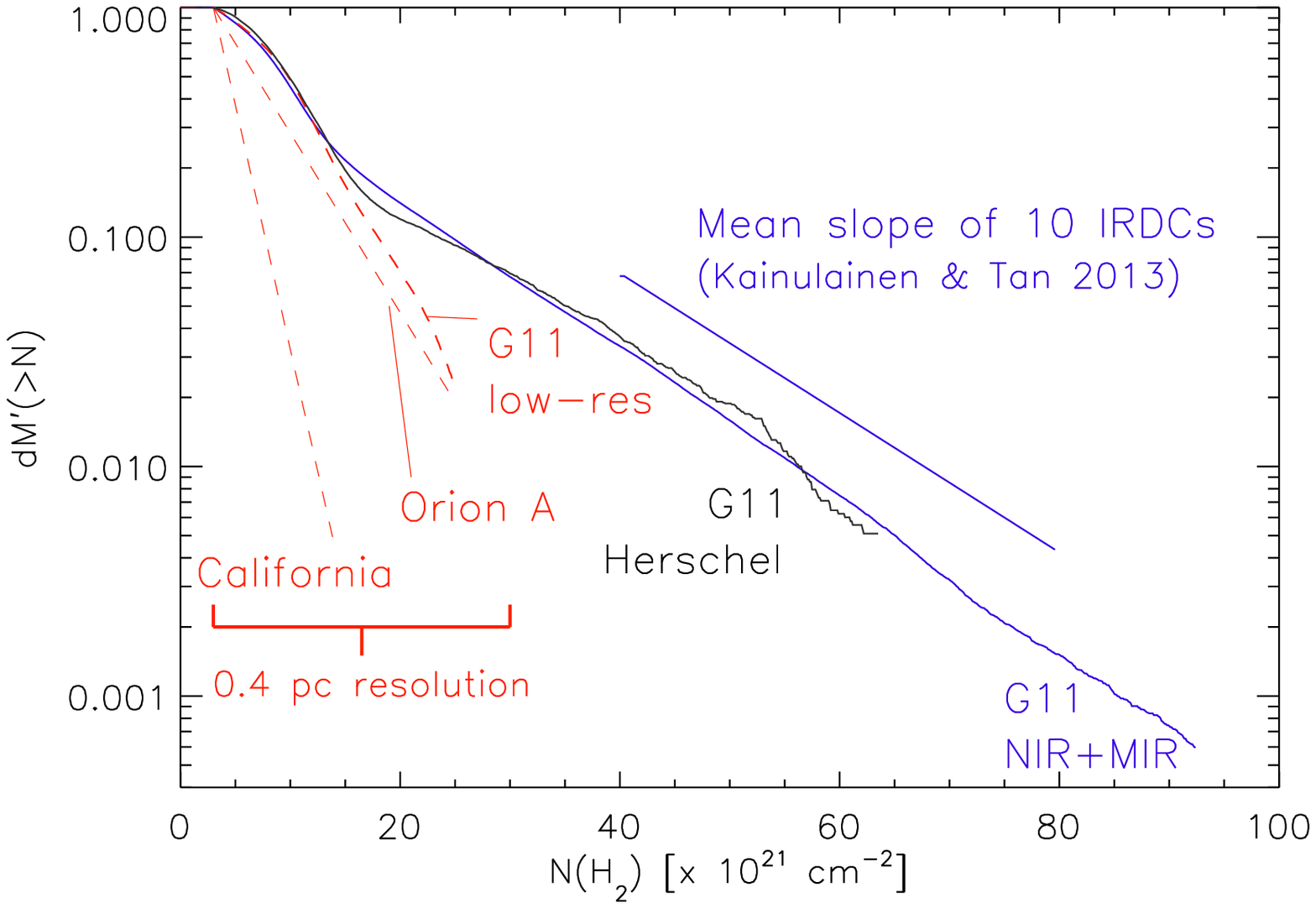}
      \caption{DGMFs of G11, derived using NIRMIR dust extinction (blue) and \emph{Herschel} (black). The red lines show the DGMFs of the California Cloud and Orion A \citep{kai09}, and of G11 with the same spatial resolution and column density range. The figure also shows a blue line indicating the slope of an exponential fit to the the mean DGMF of ten IRDCs from \citet{kai13}. The fit was done to the DGMFs  above $N(\mathrm{H}_2) \gtrsim 7 \times 10^{21}$ cm$^{-2}$. 
              }
         \label{fig:dgmfs}
   \end{figure}

We examined the distribution of mass in G11 by analyzing its \emph{dense gas mass fraction} \citep[DGMF hereafter, e.g.,][]{lom08, kai09, kai13}, which describes the cumulative mass as a function of $N(\mathrm{H}_2)$:
\begin{equation}
\mathrm{d}M' (> N) = \frac{M(> N)}{M_\mathrm{tot}},
\label{eq:dgmf}
\end{equation}
where $M(> N)$ is the mass above the column density $N$ and $M_\mathrm{tot}$ the total mass. Figure \ref{fig:dgmfs} shows the DGMFs derived from both the dust extinction and \emph{Herschel} data. They are very similar, except that the \emph{Herschel}-derived DGMF stops at lower column densities because it averages column densities over larger area than the extinction data. The DGMFs are close-to exponentials above $N(\mathrm{H}_\mathrm{2}) \gtrsim 15 \times 10^{21}$ cm$^{-2}$ and have a slight deviation from it below that. An exponential fit, $dM' \propto e^{{-\alpha N(\mathrm{H}_\mathrm{2})}}$, yields the slope of $\alpha = -0.07$. This corresponds exactly to the mean DGMF derived for ten IRDCs by \citet{kai13} and shows that G11 is a typical high-mass IRDCs by its mass distribution. 

We also compare the DGMFs of G11 to nearby, equally massive clouds Orion A and the California Cloud \citep{kai09}. Orion A is an active star-forming cloud, while the California Cloud is clearly more quiescent \citep{lad09, har13}. The DGMFs can depend on the resolution and dynamic range of the data. Therefore, we processed the $N(\mathrm{H}_2)$ map of G11 to correspond to the data of \citet{kai09}. We first truncated the G11 data to $N(\mathrm{H}_\mathrm{2}) < 25 \times 10^{21}$ cm$^{-2}$. Then, they were smoothed to $0.4$ pc resolution. The resulting DGMF (Fig. \ref{fig:dgmfs}) of G11 resembles that of Orion A. 

Recently, \citet{lad12} suggested that the DGMF has a major role in setting the star formation efficiencies (SFEs) in molecular clouds throughout the entire mass scale of star formation. This picture was based on the observations that the DGMFs of nearby star-forming clouds are clearly flatter than those of quiescent clouds \citep{kai09, lad10} and that star-forming rates of molecular clouds correlate better with the amount of dense gas in them than with their total masses \citep{lad10}. However, G11 (and other IRDCs) shows a relatively flat DGMF despite its arguably early stage of evolution and low SFE. This implies that the DGMF and SFE may not be as closely linked as earlier suggested. Indeed, in \citet{kai13b} we examined the physical parameters affecting the DGMFs in a sample of numerical turbulence simulations. We found that in addition to SFE, the average mixture of gas compression is an important (and in fact, the dominant) parameter in setting the observed DGMFs. The flat DGMF of G11 supports this picture: it appears that G11 has developed a large amount of dense gas despite its low SFE. One possible explanation for this is that the environment in which G11 resides is different from that of local clouds. This kind of galaxy-scale variations in the density statistics of molecular clouds has recently been found in the nearby M51 galaxy \citep[][]{hug13}. In M51, the spiral arm regions show higher relative amounts of dense gas than inter-arm regions. We hypothesize that the same phenomenon is seen when the DGMFs of nearby clouds are compared with G11.


\subsection{Tracing multi-scale fragmentation in G11}           
\label{subsec:fragmentation}

The high-spatial resolution and sensitivity of the extinction data allows us to study fragmentation in G11 over wide ranges of sizes and densities. In general, describing the hierarchical structure of molecular clouds is non-trivial, and there are many tools available for the purpose \citep[e.g.,][]{stu98, alv07, bur09}. Here, we aim at an analysis that describes the scale-dependency of fragmentation in G11.
We use a size-scale-based approach employed in \citet{alv07}. It entails performing a wavelet-decomposition of the $N(\mathrm{H}_2)$ data and identifying significant, single-peaked structures at different filtering scales of the decomposition. The approach is analogous to high-pass filtering, but also incorporates object recognition over multiple spatial scales. This makes the method less prone to detect spurious structures that are present in only one scale and do not correspond to real column density features \citep{kai09lada}. The algorithm results in identification of structures at scales that are $\{ 4, 8, 16, \dots   \}$ pixels in size, corresponding to $\{ 0.08, 0.17, 0.34, \dots   \}$ pc. For example, 128 significant structures are found at the smallest (0.08\,pc) spatial scale. 

\begin{figure}
   \centering
\includegraphics[bb = 0 12 480 340, clip=true, width=\columnwidth]{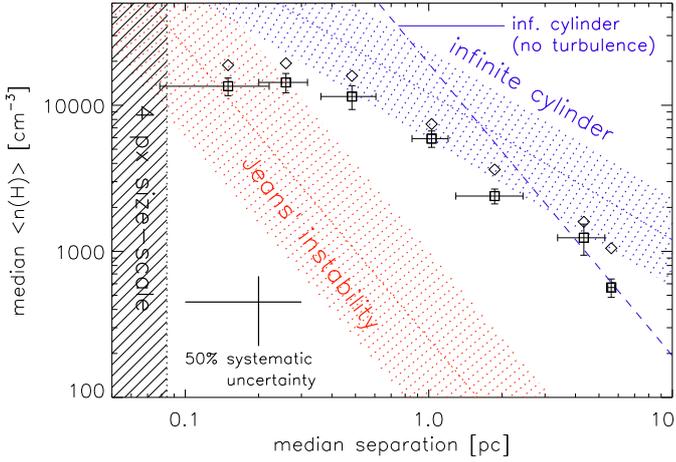}
\caption{Median mean densities ($\overline{n}_\mathrm{H}$) of significant structures at different spatial scales in G11 as a function of the separation of the structures. The squares and diamonds show the densities assuming spherical and cylindric geometries, respectively. The red dotted line shows the prediction for Jeans' instability. The blue dotted line shows the prediction for the self-gravitating cylinder, assuming the Larson's velocity dispersion-size scaling (see text). The colored areas indicate the area within a factor of two of the predictions. The blue dashed line shows the prediction for the self-gravitating cylinder in the case of only thermal support. The error-bars on data points show the inter-quartile ranges. For visibility, they are only shown for the spherical geometry. The error-bar in the lower left corner indicates 50\% systematic uncertainty.     
              }
        \label{fig:scales}
  \end{figure}

We identified structures located within the large-scale filament, and calculated their median separations and mean densities. We used both spherical and cylindric approximations in calculating volume densities. Assuming spherical geometry gives $\overline{n}_\mathrm{H} = M / (\pi (A/\pi)^{3/2} \mu m_\mathrm{H})$ where $M$ and $A$ are the mass and area, and $\mu$ and $m_\mathrm{H}$ the mean molecular weight and hydrogen mass. The cylindric geometry gives $\overline{n}_\mathrm{H} = M / (\pi b^2 h \mu m_\mathrm{H})$, where $b$ and $h$ are the minor axis of the structure and its length, respectively. The relationship of the resulting volume densities and separations is shown in Fig. \ref{fig:scales}. The figure also shows the predictions from the spherical Jeans' instability and the analogy of it in infinitely long, cylindric case \citep[e.g.,][]{cha53, inu92, jac10}. In the former model, the core separation is related to density via Jeans' length, $l_\mathrm{J} = c_\mathrm{s}(\pi/(G \overline{\rho}))^{1/2}$. In the latter model, the separation depends on the filament scale height $H = c_\mathrm{s}(4\pi G\rho_\mathrm{c})^{-1/2}$ and radius $R$. When $R >> H$, the separation is $\lambda_\mathrm{max} = 22 H$. In G11, the mean density at largest scales is $\overline{n}_\mathrm{H} \approx 10^{3}$ cm$^{-3}$ and $H \approx 0.17$ pc, assuming $T= 15$ K. The radius of the large-scale filament is $\sim$0.7 pc. Thus, it appears that we can estimate the separation scale with $\lambda_\mathrm{max} = 22 H$ \citep[as in][]{jac10}. However, this prediction is for non-turbulent case, while G11 should be mainly supported by non-thermal motions (cf., Section \ref{subsec:cd}). 
While we do not have detailed velocity data for the purpose, we assume Larson's relationship, $\sigma_\mathrm{v} = 720$ m s$^{-1} (R / 1 \mathrm{\ pc})^{0.5}$, as an estimate of line-widths in G11, which is in rough agreement with pointings in NH$_3$ \citep{pil06b}. The predicted separations are calculated by replacing the $c_\mathrm{s}$ with $\sigma_\mathrm{v}$ and are shown in Fig. \ref{fig:scales}.

The fragmentation characteristics of G11 are in agreement with the predictions for a self-gravitating cylinder at scales $l \gtrsim 0.5$ pc. At smaller scales, the relationship between density and separation appears to change, and the observations are in better agreement with Jeans' fragmentation. We hypothesize that at large scales, the fragmentation is driven by the global, filamentary nature of the cloud, while at smaller scales local fragmentation is insensitive to the large-scale geometry. We also consider that the finite nature of the G11 filament may play a role in the collapse time-scales in light of the recent predictions by \citet{pon11}.  They relate the local and global collapse time-scales by 
$(\tau_\mathrm{local} / \tau_\mathrm{global})^2 = (1 + \frac{\epsilon L^2}{L_1^2} )$, 
where $L$ and $L_1$ are the global and local size-scales, respectively, and $\epsilon$ is the relative enhancement of mass line density that serves as the seed of the collapse. It follows that the time-scale of local collapse becomes much (more than ten times) shorter than global collapse at the size-scale of $[0.2, 2]$\,pc, corresponding to choices of $\epsilon = [0.01, 0.5]$. While this range of scales is large, it shows that there indeed \emph{should be} a size-scale at which the fragmentation becomes physically dominated by local instead of global processes. Our observations of G11 suggest that this size-scale is at about $\sim 0.5$\,pc where we see a flattening in the density-separation relation.

The scale-dependency of fragmentation in G11 may relate to the velocity structure of the cloud. Recently, \citet{hac13} showed that the 10 pc long B213 filament is built up by $r \gtrsim 0.5$ pc sized filament "bundles". The bundles contain multiple velocity-coherent filaments that are further fragmented into cores.  They propose that this hierarchy may originate from scale-dependent contraction time-scales. Our results for G11 fit well in this picture. The transition in fragmentation characteristics of G11 occurs at $\sim$0.5 pc, which is the characteristic size of the velocity-coherent filaments identified by \citet{hac13}. One explanation is that at these scales the local collapse time becomes sufficiently short compared to the global collapse time, enabling small-scale fragmentation (to dense cores) unaffected by large-scale instabilities. We will investigate this further with kinematic data in a forthcoming paper (Ragan et al. in prep.).

\section{Conclusions} 
\label{sec:conclusions}

We analyze the column density distribution of the high-mass, filamentary IRDC G11.11-0.12 (G11) using combined NIR+MIR dust extinction data and \emph{Herschel} dust emission data. Our conclusions are as follows.

\begin{enumerate}

   \item The combined NIR+MIR dust extinction map reveals the column density structure of the G11 filament in a resolution of $2\arcsec$ (0.035\,pc) over the range of $N(\mathrm{H}_2) \approx 2-100 \times 10^{21}$\,cm$^{-2}$. The column densities derived from the NIR+MIR data and \emph{Herschel} data are in excellent agreement. The total mass of G11 is about $10^5$\,M$_\odot$, from which $\sim 2 \times 10^4$ M$_\odot$ is associated with the 30\,pc long filamentary structure. 
   
   \item The dense gas mass fraction (DGMF) of G11 is typical for high-mass IRDCs, showing that it harbors a relatively large amount of dense gas, comparable to active nearby star-forming clouds, e.g., Orion A. The linear mass density of the large-scale filament in G11 is high, $M_\mathrm{l} \approx 600$ M$_\odot$ pc$^{-1}$, also similar to the Orion filament ($M_\mathrm{l} \sim 325$ M$_\odot$ pc$^{-1}$). However, the star-forming activity of G11 is arguably much lower than that of the Orion filament. This suggests that the amount of dense gas in molecular clouds is not directly connected to their star-forming efficiencies \citep[c.f.,][]{lad10}, but is rather set by other parameters/processes, possibly prior to the most active star-forming phase of the clouds. One explanation that can account for this is the average level of gas compression in molecular clouds \citep[c.f.,][]{kai13b} that depends on the galaxy-scale environment. This picture is supported by the environment-dependence of the molecular clouds' density statistics in M51 \citep[e.g.,][]{hug13}.
      
   \item We find that the structure of G11 is in agreement with hierarchical fragmentation of a self-gravitating cylinder at sizes larger than $r \gtrsim 0.5$\,pc. At smaller scales, the fragmentation more closely resembles Jeans' fragmentation. This suggests that the fragmentation at large scales is dominated by the collapse of the natal filament, while at small-scales the filamentary origin does not play a role, but fragmentation depends on local properties only. This hypothesis is in agreement with scale-dependent collapse time-scales derived for finite filaments \citep{pon11}.  
   
\end{enumerate}


\begin{acknowledgements}
The authors are grateful to Fabian Heitsch, Henrik Beuther, and Alvaro Hacar for helpful discussions.
The work of JK, SR, and AS was supported by the Deutsche Forschungsgemeinschaft priority program 1573 ("Physics of the Interstellar Medium").
\end{acknowledgements}


\pagebreak


\appendix

\section{Comparison of dust extinction and \emph{Herschel}-derived column densities}

\subsection{Dust opacity law}

\begin{figure*}
   \centering
\includegraphics[width=\textwidth]{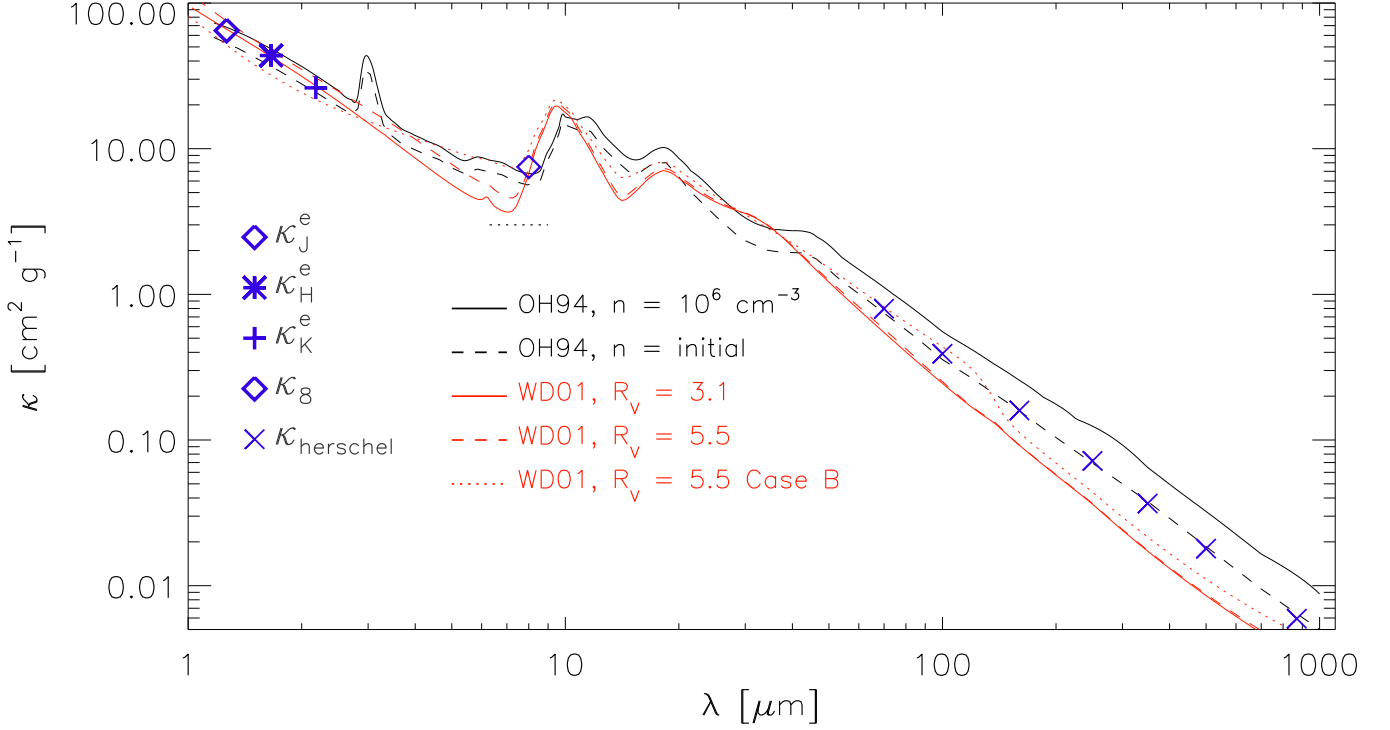}
      \caption{Dust opacity-law adopted for the work. The five lines show different plausible dust opacity models from \citet{oss94} and \citet{wei01}. The symbols show the adopted dust opacities in NIR and 8 $\mu$m \citep[cf.,][]{kai13} and in \emph{Herschel} wavebands. The adopted values represent the mean values in the range set by the five shown models. 
              }
         \label{fig:opacities}
   \end{figure*}

Figure \ref{fig:opacities} shows the absolute dust opacity law adopted in transforming NIR+MIR optical depths and sub-mm dust emission data to column densities. The figure shows five plausible dust opacity models. These were the \citet{oss94} models with moderate dust coagulation and without dust coagulation, and \citet{wei01} models with different $R_\mathrm{V}$ values. The effective dust opacities used in NIR and 8 $\mu$m are close to the mean values of these models. Therefore, we chose in this work to use a corresponding definition to set the sub-mm opacities. The chosen values are shown in the figure and they are nominally closest to the \citet{oss94} model without dust coagulation. 

\subsection{Effect of variable background to the \emph{Herschel}-derived temperatures and column densities}

\begin{figure*}
   \centering
\includegraphics[bb = 20 200 600 650, clip=true, width=\columnwidth]{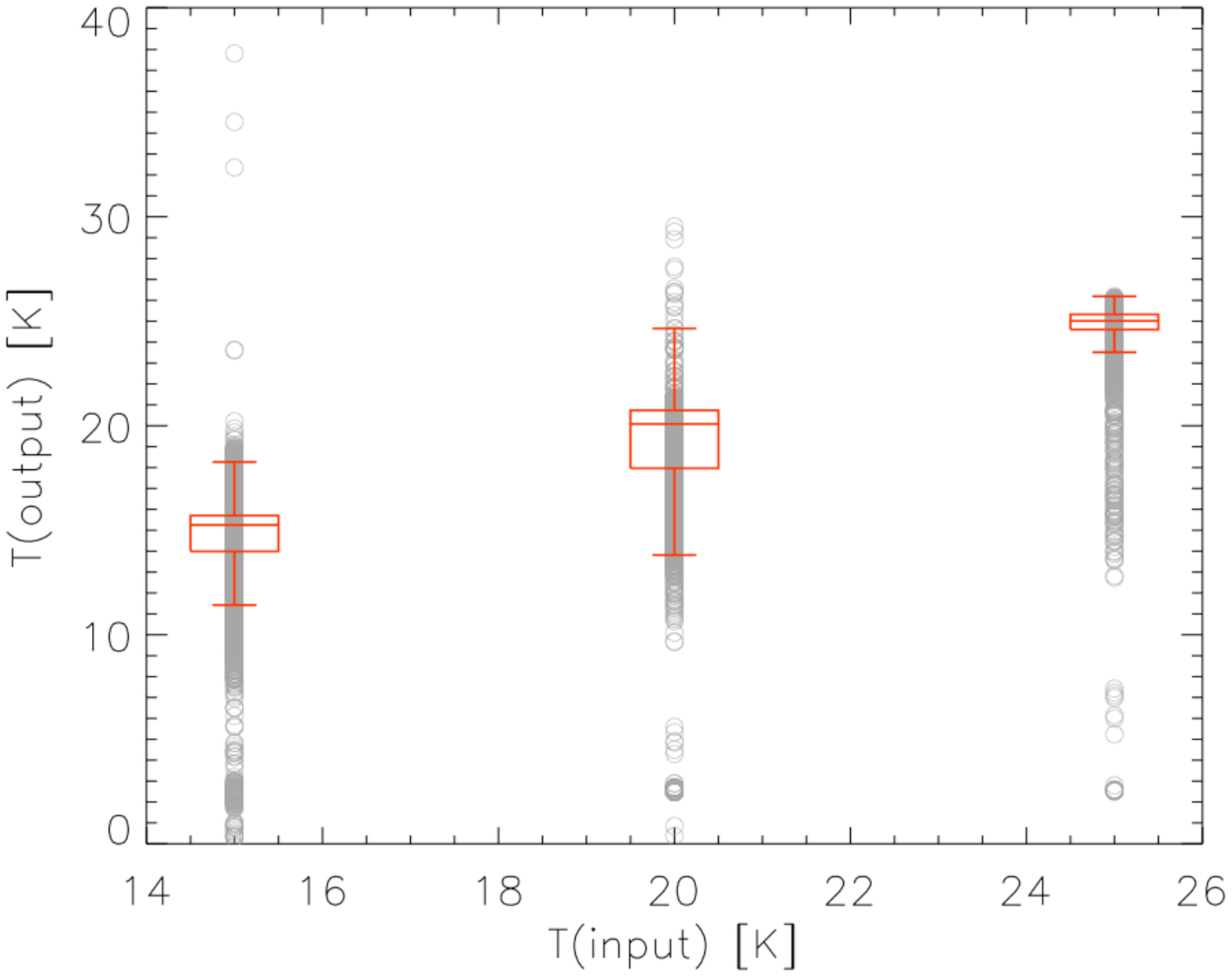}
\includegraphics[bb = 20 200 600 650, clip=true, width=\columnwidth]{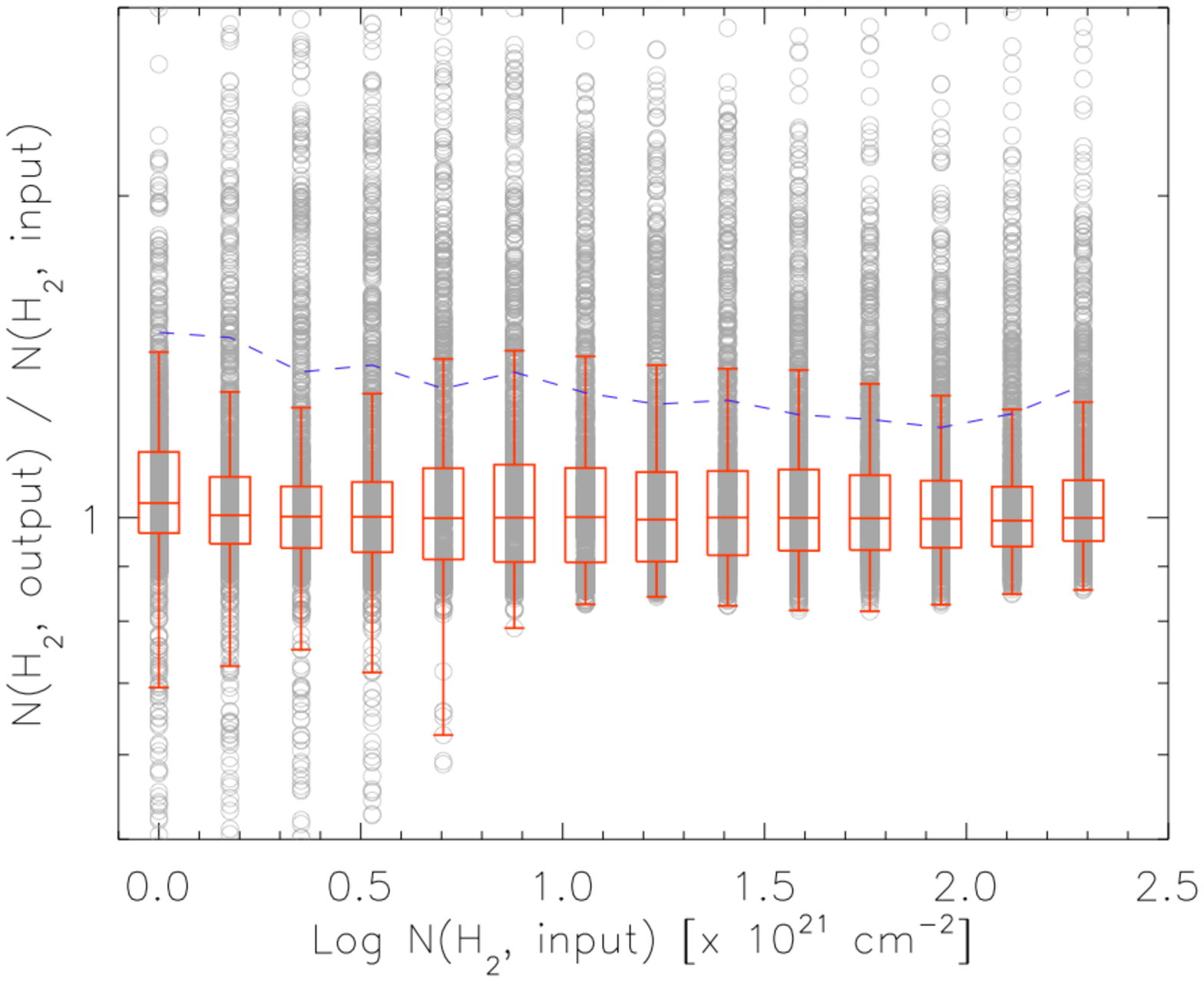}
      \caption{Effect of the variable background emission to the \emph{Herschel}-derived temperatures and column densities. The panels show results of a Monte Carlo simulation, in which normally-distributed random variations were imposed on a large sample model SEDs. \emph{Left: }The output temperature for three values of input temperatures, $\{15, 20, 25\}$ K. The repetitions of the simulation are shown with light-grey circles. The red boxes indicate inter-quartile ranges, with a horizontal line at the mean value, and the whiskers show ranges that extend to 1.5 times the inter-quartile range. \emph{Right: }The same for the ratio of input to output column densities as a function of input column density. The blue dashed line shows the standard deviation, which is approximately 30\% irrespective of the original column density.
              }
         \label{fig:bg_sim}
   \end{figure*}

We performed a Monte Carlo simulation to examine the effects of variable background emission to the column densities derived using \emph{Herschel} dust emission data and the fitting technique described by \citet{Launhardt2013}. Figure \ref{fig:bg_sim} summarizes the results. The results are presented and discussed in Section \ref{subsec:comparison}.

\end{document}